\documentclass[prb,preprint]{revtex4-1} 
\usepackage{amsmath}  
\usepackage{amsfonts} 
\usepackage{graphicx} 
\begin{document}

\title{Effects of exoplanetary gravity on human locomotor ability}
\author{Nikola Poljak, Dora Klindzic and Mateo Kruljac}
%\email{npoljak@phy.hr}
\affiliation{Department of Physics, Faculty of Science, University of Zagreb, Croatia}
\date{\today}

\begin{abstract}
At some point in the future, if mankind hopes to settle planets outside the Solar System, it will be crucial to determine the range of planetary conditions under which human beings could survive and function. In this article, we apply physical considerations to future exoplanetary biology to determine the limitations which gravity imposes on several systems governing the human body. Initially, we examine the ultimate limits at which the human skeleton breaks and muscles become unable to lift the body from the ground. We also produce a new model for the energetic expenditure of walking, by modelling the leg as an inverted pendulum. Both approaches conclude that, with rigorous training, humans could perform normal locomotion at gravity no higher than 4\,$g_{Earth}$.\\\\
To be published in \bf{The Physics Teacher}.
\end{abstract}

\maketitle 

\section{Introduction} 

With the discovery of many new potentially habitable exoplanets, one needs to consider which ranges of planets' physical parameters are suitable for immediate human settlement. Aside from the average temperature, insolation, pressure, atmospheric composition, etc., all of which can be solved with spacesuits, the basic parameter of a planet is its \textbf{surface constant of gravity}, which will determine if a person can stand upright and move safely and at a reasonable pace from one place to another.

Studies of animal sizes \cite{hokkanen} have already determined many physical limits the animals can reach in conditions existing on Earth. In this text, we aim to take another approach, fixing the size of the animal, in this case a human, and determining the range of gravitational accelerations $g$ in which it can stand and move. To do so, we will look at the largest $g$ in which our skeleton still won't fail and in which our muscles can still perform the basic movements of standing up and walking.

\section{Methods and results}

\subsection{Bone failure}

\noindent Let us, for the sake of simplicity, imagine the whole human weight being supported by a single upright bone, in this case representing the entire skeleton. The weight $Mg$ of the entire human mass acts on the bone with a diameter $D$ and a cross section $A$ and produces an axial compression equal to:
\begin{eqnarray}
\sigma = \frac{Mg}{A}\propto \frac{Mg}{D^2}\propto gD \,,
\end{eqnarray}
where we used the fact that the cross section of the bone is proportional to the square of its dimension and the mass of the human to its cube. Since we want to obtain a numerical result, it is not enough to deal with proportionalities. Measurement data \cite{alexander79,fung81} shows that for an average 50\,kg mammal, which we can take as a first approximation of a human, the cross sectional area $A$ of a tibia (which we consider to be at least under the same pressure as the rest of the skeleton) equals $2.7 \cdot 10^{-4} \textrm{m}^2$ and the compressive strength $\sigma$ of the bone is about 170MPa, giving the maximum gravity the human can support as:

\begin{eqnarray}
g_{\textrm{max}} = \frac{\sigma A}{M} = 918 \,\textrm{m}/\textrm{s}^2 \,.
\end{eqnarray}

This is indeed a large number, corresponding to more than 90 times the Earth gravity! However, this maximum value needs to be reduced since it considers only static compressive stress on the bone. Once we start moving, the dynamic stress takes over due to bending of the bones subject to gravitational torques. It has been shown experimentally \cite{alexander79,biewener83} that the total stress increases approximately by a factor of 10 during normal running, thus reducing the maximum gravity to $\approx 10$ times the Earth gravity. The same studies suggest that a factor of 10 might be too large, however, as these were conducted on larger animals (such as cows) they can not be reliably extrapolated to human sized mammals. Further, an increased compressive force acts on the bones due to a ground reaction force during running or jumping. Since we accelerate vertically when running, the normal force on the bones of the legs is greater than the force due to gravity. However, studies \cite{Cross} show that this increases the stress on the bone by at most a factor of 2, so it doesn't modify our result.

\subsection{Muscle strength} \label{snaga}

\begin{figure}[h!]
  \centering
  \includegraphics[width=0.5\textwidth]{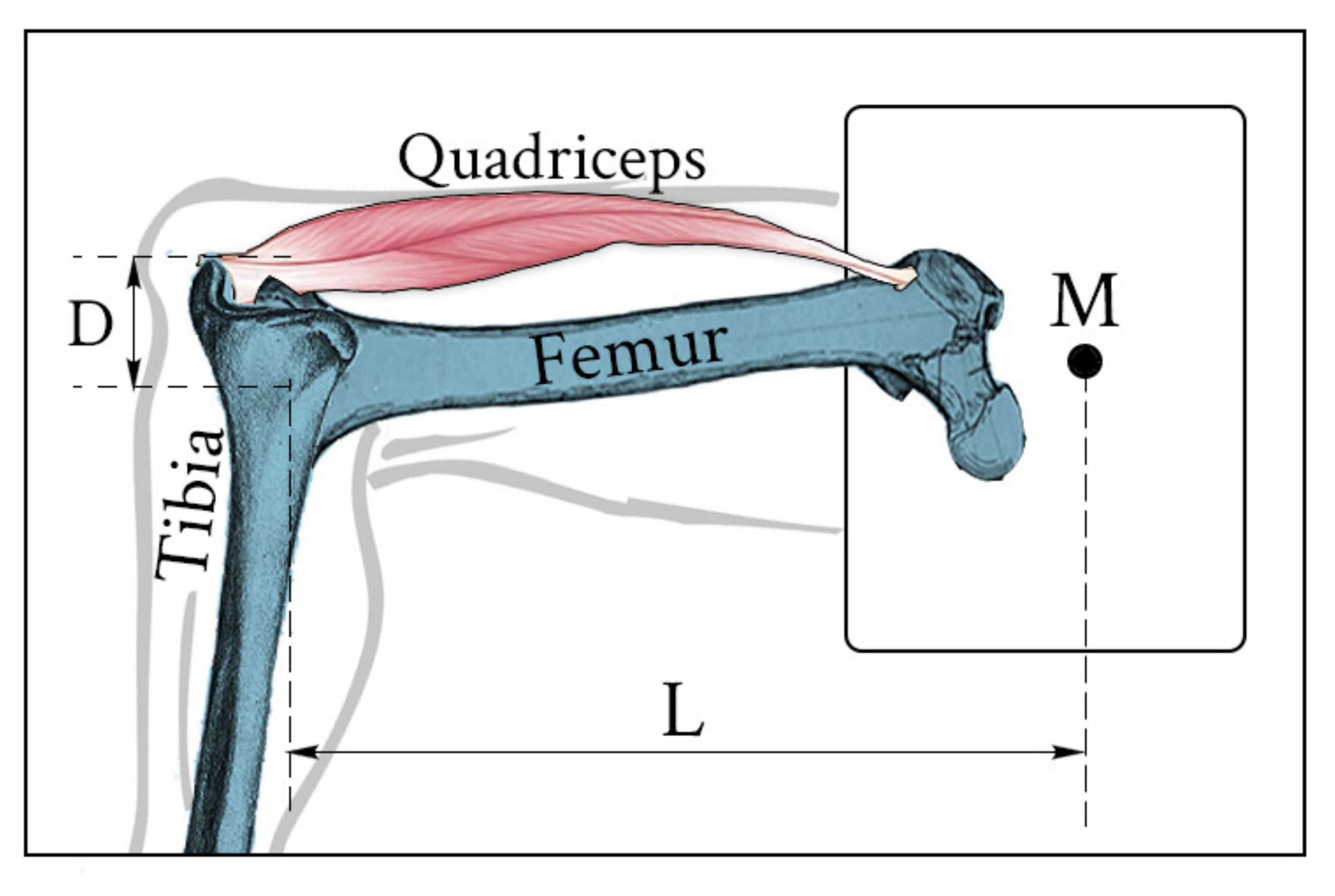}
\caption{A representation of the human leg. The human has a mass $M$. The distances denoted in the image are discussed in the text.}
\label{figure1}
\end{figure}

\noindent A criterion for muscle strength will be the ability of the human to get up while seated or lying down. A visual representation of the problem is given in Fig \ref{figure1}. 

We consider the entire mass $M$ of the human to be located at their center of gravity. The quadriceps connects the massless femur and tibia and is responsible for getting up.  In order to do so, the force with which the quadriceps must pull has to be at least:
\begin{eqnarray}
F \geq \frac{MgL}{2D}\,,
\end{eqnarray}
in order to at least balance the gravitational torque on mass $M$. In this expression, $D$ is the torque arm of the muscle force and $L$ is the torque arm of the gravitational force. The factor 2 arises since humans have 2 legs and each needs to lift only half of the total mass. Another way to express the maximum force the quadriceps produces is with the help of the maximum isometric stress $\sigma_m$ muscles can produce:
\begin{eqnarray}
F = \sigma_mA_m=\sigma_m\frac{M_m}{\rho_mL_m}\,,
\end{eqnarray}
where the indices $m$ denote the muscle and $A$ is the muscle cross section, expressed in the second equation with the help of muscle mass $M$, its density $\rho$ and and its length $L$. Once again, the values for these parameters are known for a mammal of 50\,kg \cite{alexander79,alexander81}. Finally, the torque arm $L$ is longer than the muscle length and approximated by $L=1.5L_f$, where $L_f$ is the femur length. Plugging in all the numerical values, we obtain the limiting value of $g$:

\begin{eqnarray}
g_\textrm{max} = \frac{2\sigma_mM_mD}{\frac{3}{2}M\rho_mL_mL_f} = 10.7 \textrm{m}/\textrm{s}^2 \,.
\end{eqnarray}

Note that this result is reproduced for a ``worst-case scenario'' - it is very difficult to get up from a sitting position based only on muscle strength without bending forward or pushing your feet slightly backwards. Still, we think this is an extraordinary result, showing how well our muscle system is adapted to life on Earth! It would appear that living on planets with increased gravity would be very difficult due to our relatively weak muscles. However, we could either try to increase our muscle strength or move around with the help of technology. Evidence from Earth suggests that increase in muscle strength is possible within some limits. Looking at isometric squat standards for 50\,kg men \cite{sqt}, one can see that an average person can lift about 36\,kg, while an elite athlete can lift 145\,kg, which is an increase by a factor of $\approx$4. Of course, we also use our arms and other muscles when getting up, which we have to factor in and choose as an extra 20-50\%. We choose to take the lower limit to this estimate since we believe that an increase of factor of 4 that we included earlier could be a slight overestimate since not everyone can become an elite athlete. Thus, one could assume that with rigorous training, we could get up at gravity values of $\approx 10.7 \cdot 4\cdot 1.2\: \textrm{m}/\textrm{s}^2 \approx 5 g_\textrm{Earth}$. 

One could argue that we could have used the world squat record for a 50\,kg man, but that's a result only one person can achieve, while there are quite a few elite athletes. The latter is far more useful information if we wish to colonize a planet. Since the quoted strength increase factor of $\approx$4 is the same for all weight classes \cite{sqt}, 5\,$g_{Earth}$ should be the maximum gravity for them as well.

We can see that increased gravity induces stress on the muscular system much more so than on the skeletal system. This was to be expected, since we all know it is relatively easier to get strong than to break a bone. The limit on the surface gravitational constant therefore arises mostly from the ability to get up from the floor using your muscles. However, assuming that vehicle-assisted transportation is unacceptable for long-term settlement on planets, we must also examine the energetic expenditure of walking.

\subsection{Locomotion}

\begin{figure}[ht]
\centering
\includegraphics[width=0.5\textwidth]{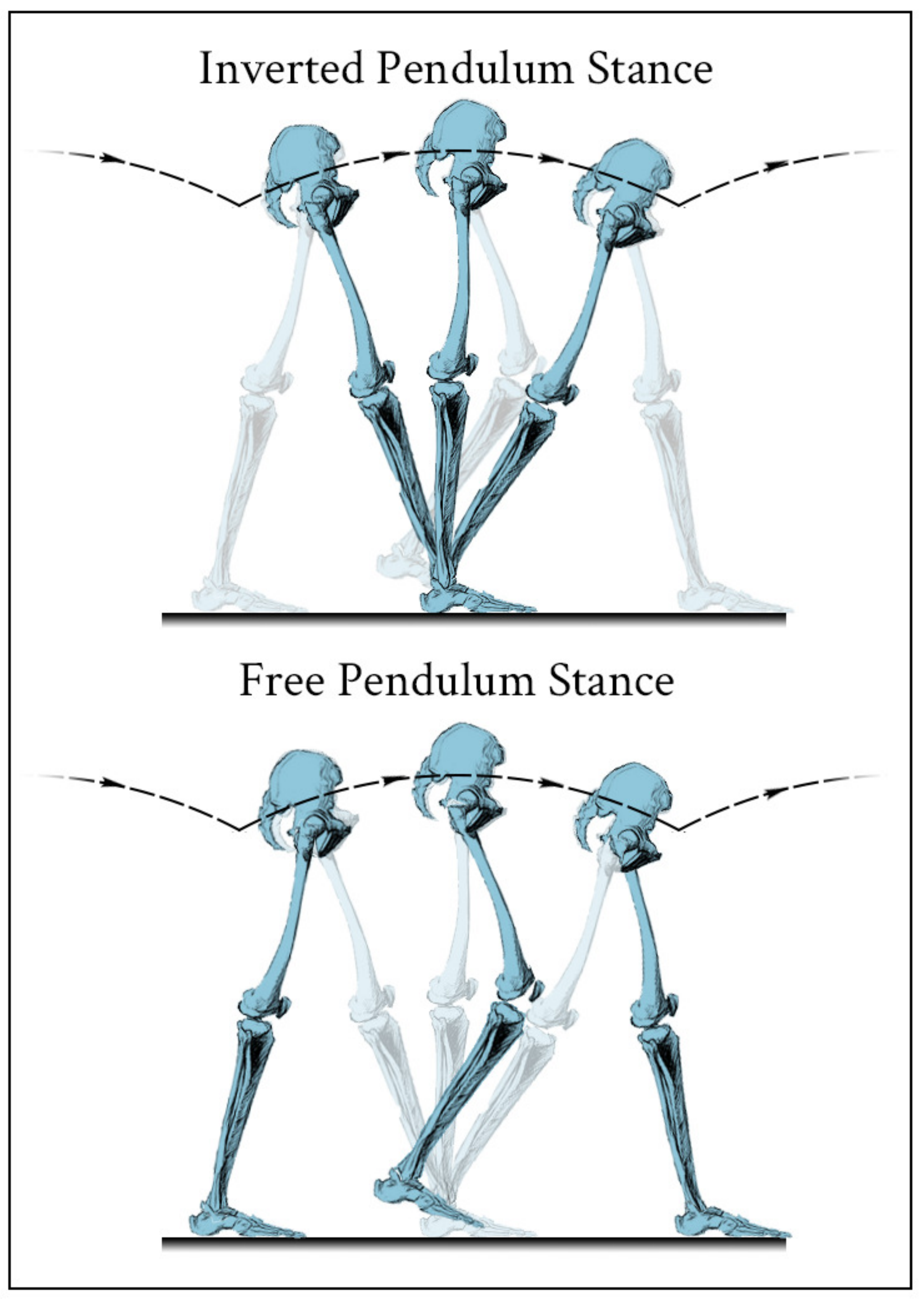}
\caption{Whilst walking, one leg behaves as an inverted pendulum, and the other as a swinging pendulum.}\label{fig:gait}
\end{figure}

\noindent Walking can be accurately represented by the '\textbf{inverted pendulum gait}', as in figure \ref{fig:gait}. The leg that supports the weight is stiff and behaves like an inverted pendulum, with the body's center of mass on top. The free leg swings forward as a free pendulum, although in reality, it's not a completely conservative system - we know we have to use energy to swing it. In an ideal scenario, though, once an organism started walking in such a stance, it would not require any energy to maintain the walk. Therefore, walking in the inverted pendulum gait is the most energy-efficient means of limb-assisted locomotion, and since it appears in every land animal on Earth, we can assume that it will be present in extraterrestrial life, too.

Now, we may consider that most of the work is spent on making the center of mass (CM) 'bob' up and down \cite{romeo}. In that case, the CM repeatedly oscillates between maximum kinetic energy, at the bottom of the movement arch, and maximum gravitational potential energy at the top. The ratio of those two energies is proportional to

\begin{equation}\label{eqn:froude}
F = \frac{v^2}{gL},
\end{equation}
where $v$ is the traveling speed, $g$ is the surface gravitational constant and $L$ is the limb length. The number $F$ is called the \textbf{Froude number}, and studies have conclusively found that animals with nearly the same $F$ exhibit the same gait \cite{evolution,gazette}. That is, the ratio of kinetic and potential energy of the CM determines if an animal will walk, run, trot or gallop at a certain speed, regardless of species.

How does this apply to the walk of bipedals, quadrupedals, or other creatures with an even number of legs? At the high point of the CM 'bobbing' at each pair of legs, there is a centrifugal force lifting it upwards, and at the same time, a gravitational force downwards. If the centrifugal force exceeds it, it will become impossible to keep the foot on the ground. The condition for this is:
\begin{equation}
\frac{mv^2}{L} < mg, \quad \text{or} \quad F<1.
\end{equation}
When the Froude number exceeds 1, walking becomes impossible. In reality, nature never pushes physics to the limits, so humans and animals consistently adopt a trot or run at around F = 0.5. A study \cite{walk-run} has shown that the transition from walking to running occurs at F$\approx$0.5 regardless of gravity, by simulating a low-g environment and measuring the speed at which the transition occurs. This should work for every surface and every planet.

\begin{figure}[h!t!]
\centering
\includegraphics[width=0.5\textwidth]{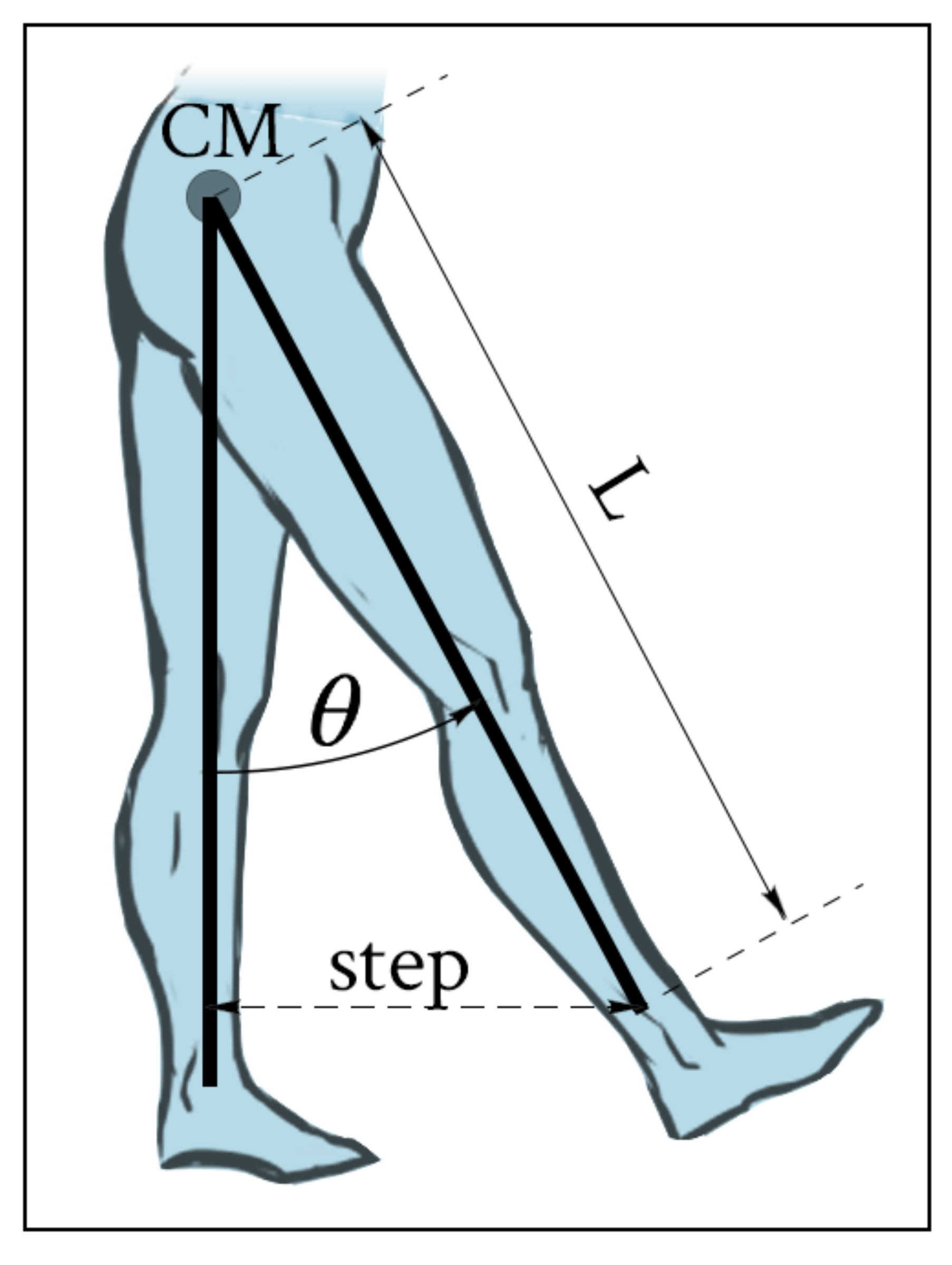}
\caption{A sketch of our pendulum model. The legs have length L and are separated by an angle $\theta$ before each step.}\label{fig:pendulum}
\end{figure}

Now that we have the link between movement speed and limb length, let's see what we can say about the energetics of walking. The minimum work required to 'bob' the CM upwards for a step that has the legs separated by an angle $\theta$ (figure \ref{fig:pendulum}) is equal to the difference of gravitational potential energies of the feet,
\begin{equation}
W_s = mgL \, (1-\cos \theta ).
\end{equation}
To avoid dealing with angles and make a connection with the Froude number, we will use the step length $s$ as a parameter, so we identify $\sin \theta = s/L$, and $s = v \, t_s$, where $t_s$ is the time required for one step. Since the most energy-efficient way for the swinging leg to move is like a stiff pendulum, we can take the step time to be one quarter of the period of such a pendulum.

Using equation (\ref{eqn:froude}) and the natural period of the human leg, $4t_s = T = 2\pi \sqrt{2L/3g}$ \cite{sprott}, we get:
\begin{equation}\label{eqn:work}
W_s = mgL \, \left( 1-\sqrt{1 - \frac{\pi^2}{6}F} \right).
\end{equation}

Let's put this to the test. According to medical information \cite{wolfram} used by the Wolfram Alpha computational engine, a woman of 65 kilograms and 20 years of age, briskly walking at 1.4 m/s, makes 100 steps per minute and spends 15 kJ of energy. That is roughly 150 Joules per step. Her Froude number should be around 0.3. By plugging all this into equation (\ref{eqn:work}), we can estimate her leg size as 0.8 meters. This turns out to agree with the average leg length for the female population of this age \cite{noge}, as well as the popular '45\% of total height' estimate.

Notice that for equation (\ref{eqn:work}) to hold, the expression under the square root has to be positive, i.e. $F\leq 0.6$. As  mentioned before, it's been shown that the transition from walking to running occurs at $F\approx 0.5$ and the pendulum model does not apply for larger values of F anyway. Thus, we can assume our model is accurate within the order of magnitude for all creatures with the inverted pendulum gait. The energetic contribution of the swinging leg to energy expenditure is a few Joules at best, due to the leg's small mass.

Let's try to find the maximum gravity at which we could make a step. For that purpose, we'll look at the record set by a strongman, who walked 5 steps with a 649\,kg log on his back \cite{mountain}. We used the world record because there are no official elite athlete results for log carrying, as we've found for squatting. Regardless, it is reasonable to assume this result is close enough to an elite athlete so that any error in setting him as our benchmark is probably negligible. We'll use equation (\ref{eqn:work}) to find the gravity at which the strongmans ``free'' walking is the same as walking with that log here on Earth. Now, he was walking very slowly, meaning $v^2\ll 1$, i.e. $F\ll 1$. We will assume that  Froude numbers on both planets are small and equal and use the Taylor expansion $\sqrt{1-x}\approx 1-\frac{1}{2}x$ in equation (\ref{eqn:work}), which then reduces to:
\begin{equation}
(M_{man}+M_{log})g_{Earth}=M_{man}g_{max}
\end{equation}
which, for the strongmans mass of 179\,kg, gives $g_{max}\approx 4.6\:g_{Earth}$.

Comparing this result to the former result in subsection \ref{snaga}, we find the models give a very similar result for the maximum bearable gravity.

\section{Discussion}

Even though we found some limits on the maximal surface gravitational constant on acceptable planets, there are other considerations that have to be taken into account. For example, since increased gravity would be strongly pulling our blood down to our legs, the heart would have to work harder to pump it up into the brain. On the other hand, once we're subject to low gravity, blood rushes from our legs into the face. To adapt, the heart would need to work differently and regulate the blood pressure. Following this line of reasoning, we can conclude that subject to high gravity, blood goes from the chest region into the legs, resulting in larger blood volume and higher blood pressure. Because blood cells are more easily destroyed than created, the body's cardiovascular system should adapt sooner to low than to high gravity. Until enough blood cells are created in high gravity, we could feel weak, like after donating blood, in addition to having trouble standing and walking. 

We also know there are health risks due to low \cite{ltlak} and high \cite{vtlak} blood pressure, meaning high-g planets could cause damage to the heart, arteries, kidneys and the brain as well as leave us with dizziness, nausea and fatigue. Studies have shown that the average human body could not withstand gravity greater than 5\,$g_E$ without passing out \cite{wiki-gloc}, because the heart couldn't pump enough blood into the head, so we'll assume that 4\,$g_E$ is the maximum gravity the human body can withstand in the long run. Other health risks in a high-g environment could be as simple as falling down, since our weight would be substantially increased, but we don't consider those in detail here. In conclusion, all of the systems governing the human body we examined give the maximum gravity of about 5\,$g_E$. This should only be considered as an upper limit which could only be achieved by a handful of astronauts. It is difficult to claim what the exact upper limit for an average person could be, but we can estimate that a number above 3.5\,$g_E$ would be reasonable.

\begin{figure}[h!t!]
\centering
\includegraphics[width=0.7\textwidth]{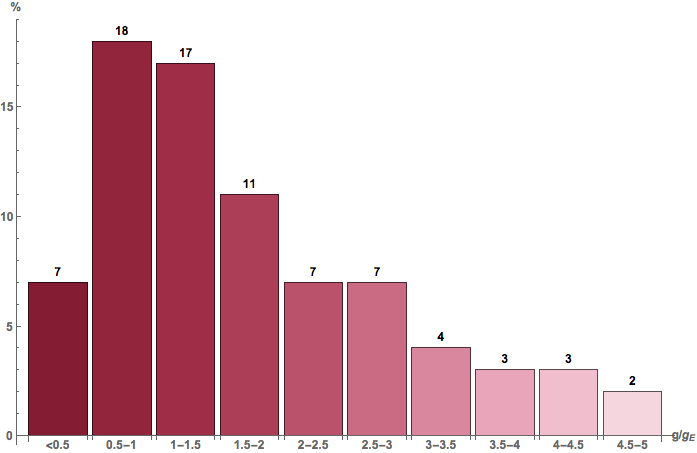}
\caption{Distribution of the gravitational constant $g$ among confirmed exoplanets \cite{archive}. The values displayed are rounded-off percentages of the total number of sub-5\,$g_E$ exoplanets made up by exoplanets in a given range of surface gravity.}\label{fig:histogram}
\end{figure}

Now that we have found 5\,$g_E$ as the upper limit of the surface gravitational constant on acceptable planets, we can look at real data \cite{archive} and find out how many exoplanets satisfy our condition. Out of 3605 confirmed exoplanets (as of January 2018), 594 have known radii and masses, which are needed to determine $g$. A short calculation shows that 469 of these fit our criterion. A chart of the distribution of $g$ is given on figure \ref{fig:histogram}. We can easily notice the peak in the percentage of planets in the gravitational range from 0.5\,$g_E$ to 1.5\,$g_E$, which represent the planets we could relatively easily adjust to. 

It should be noted that planets with gravity less than 5\,$g_E$ are in fact very low-$g$ planets when compared to 125 planets discovered with gravity stronger than 50\,$g_E$ (some of them even go as high as 200\,$g_E$). This shows that we can classify humans as low-$g$ organisms on the galactic scale and predict most exoplanetary life we may encounter to far outweigh and outgrow us \cite{simpson}. Fortunately, the majority of discovered exoplanets fit in the low sub-5\,$g_E$ group, which means a greater chance of colonization and, perhaps, meeting alien life forms which are in some ways similar to us.

\section{Conclusion}

In this study we show that the human musculoskeletal system is well-suited to Earth gravity, and would consequently have difficulty functioning under increased gravitational force. However, physical training effectively quadruples our available strength and enables us to raise the limit up to 3-4\,$g_{Earth}$. Our model of walking confirms this limit, with a strongman as our extreme example. Further it seems  that, from the human perspective, the range of habitable planets is much narrower than the range of potentially life-supporting planets in the universe. This restriction arises primarily from gravity, as the only environmental factor which we cannot manipulate. Being aware of this enables us to make our search for potential future exoplanetary colonies more precise.


\begin{thebibliography}{99}
\bibitem{hokkanen} Hokkanen, J.E.I.; ``The Size of the Largest Land Animal'', Department of Theoretical Physics, University of Helsinki, 1985
\bibitem{alexander79} R.McN. Alexander, A.S. Jayes, G.M.O. Maloiy, E.M. Wathuta; ``Allometry of the limb bones of mammals from shrews (Sorex) to elephant (Loxodonta)'', J. Zool., Lond. 189, 305 (1979)
\bibitem{fung81} Y.C. Fung; ``Biomechanics: Mechanical Properties of Living Tissues'', New York:  Springer-Verlag, 1981
\bibitem{biewener83} A.A. Biewener, J. Thomason, L.E. Lenyon; ``Allometry of quadrupetal locomotion: the scaling of duty factor, bone curvature and limb orientation to body size'', J. Zool., Lond. 201, 67 (1983)
\bibitem{Cross} R. Cross, ``Standing, walking, running, and jumping on a force plate'', American Journal of Physics 67, 304 (1999)
\bibitem{alexander81} R.McN. Alexander, A.S. Jayes, G.M.O. Maloiy, E.M. Wathuta; ``Allometry of the leg muscle of mammals'', J. Zool., Lond. 194, 539 (1981)
\bibitem{sqt} ExRx.net database, \href{http://www.exrx.net/Testing/WeightLifting/SquatStandards.html}{Squat Standards}, last edited April 2012
\bibitem{romeo} Romeo, F.; ``\href{http://arxiv.org/ftp/q-bio/papers/0609/0609007.pdf}{A Simple Model of Energy Expenditure in Human Locomotion}'', Physics Department “E. R. Caianiello”, University of Salerno
\bibitem{evolution} M.V. Rayner, Jeremy; ``Evolution on Planet Earth'', Chapter 'Gravity, the atmosphere and the evolution of animal locomotion', Appendix 10.1, page 180
\bibitem{gazette} R. McNeill, Alexander; ``Walking and running'', The Mathematical Gazette, Vol. 80, No. 488 (Jul., 1996), pp. 262-266
\bibitem{walk-run} R. Kram, A. Domingo, D. P. Ferris;
``\href{http://spot.colorado.edu/~kram/walk-run.pdf}{Effect of reduced gravity on the preferred walk-run transition speed}'',The Journal of Experimental Biology 200, 821-6 (1997)
\bibitem{sprott} Sprott, J.C.; ``\href{http://sprott.physics.wisc.edu/technote/walkrun.htm}{Energetics of walking and running}'', University of Wisconsin - Madison
\bibitem{wolfram} Wolfram Alpha computational knowledge engine, query ``\href{http://www.wolframalpha.com/input/?i=walking}{Walking}'', based on CDC standards
\bibitem{noge} Fredriks, van Buuren, et al.; ``\href{http://adc.bmj.com/content/90/8/807.full.pdf}{Nationwide age references for sitting height, leg
length, and sitting height/height ratio}'', Arch Dis Child
2005
\bibitem{mountain} The World's Strongest Men archive, \href{http://www.theworldsstrongestman.com/athletes/hafthor-bjornsson/}{Hafthor Bjornsson}, last renewed January 2016
\bibitem{ltlak} American Heart Association; Condition: ``\href{http://www.heart.org/HEARTORG/Conditions/HighBloodPressure/AboutHighBloodPressure/Low-Blood-Pressure_UCM_301785_Article.jsp#.Vn8lZvnhBD8}{Low blood pressure}'', last renewed April 2014

\bibitem{vtlak} Mayoclinic; Diseases and Conditions: ``\href{http://www.mayoclinic.org/diseases-conditions/high-blood-pressure/in-depth/high-blood-pressure/art-20045868}{High blood pressure problems}'', last renewed February 2014

\bibitem{wiki-gloc} Wikipedia; ``\href{https://en.wikipedia.org/wiki/G-LOC}{G-force induced loss of consciousness (G-LOC)}'', last modified November 2015
\bibitem{archive} \href{http://exoplanetarchive.ipac.caltech.edu/index.html}{NASA Exoplanet Archive}, a service of NASA Exoplanet Science Institute, last renewed January 2016
\bibitem{simpson} Simpson, F.; ``The Nature of Inhabited Planets and their Inhabitants'', University of Barcelona (UB-IEEC), Spain, dated: March 2015
\end{thebibliography}
\end{document}